\documentclass[prl,aps,twocolumn,superscriptaddress]{revtex4}

\usepackage[dvips]{graphicx}
\usepackage{amssymb,amsfonts,amsmath}
\usepackage{color}
\usepackage{ulem}

\newcommand{\rv}{{\mathbf r}}
\newcommand{\ov}{{\boldsymbol{\omega}}}

\newcommand{\J}{{\bf J}}
\newcommand{\Jv}{{\bf J}}

\newcommand{\vel}{{\bf v}}

\begin{document}

\title{Nonequilibrium phase behaviour from minimization of free power
  dissipation}

\author{Philip Krinninger}
\affiliation{Theoretische Physik II, Physikalisches Institut, 
  Universit{\"a}t Bayreuth, D-95440 Bayreuth, Germany}

\author{Matthias Schmidt}
\affiliation{Theoretische Physik II, Physikalisches Institut, 
  Universit{\"a}t Bayreuth, D-95440 Bayreuth, Germany}

\author{Joseph M. Brader}
\affiliation{Soft Matter Theory, 
  University of Fribourg, CH-1700 Fribourg, Switzerland}

\date{23 May 2016, revised version: 11 August 2016,
second revision: 23 September 2016}

\begin{abstract}
We develop a general theory for describing phase coexistence between
nonequilibrium steady states in Brownian systems, based on power
functional theory (M. Schmidt and J.M. Brader, J.~Chem.~Phys.~{\bf 138}, 214101 (2013)). We
apply the framework to the special case of fluid-fluid phase
separation of active soft sphere swimmers. The central object of the
theory, the dissipated free power, is calculated via computer
simulations and compared to a simple analytical approximation. The
theory describes well the simulation data and predicts
motility-induced phase separation due to avoidance of dissipative
clusters.
\end{abstract}

\maketitle

Phase transitions in soft matter occur both in equilibrium and in
nonequilibrium situations.  Examples of the latter type include the
glass transition \cite{weeks}, various types of shear-banding
instabilities observed in colloidal suspensions \cite{schall,dhont},
shear-induced demixing in semidilute polymeric solutions \cite{onuki},
and motility-induced phase separation in assemblies of active
particles \cite{phaseSeparationPapers,CatesTailleurReview}. In
contrast to phase transitions in equilibrium, which obey the
statistical mechanics of Boltzmann and Gibbs, very little is known
about general properties of transitions between out-of-equilibrium
states. A corresponding universal framework for describing
nonequilibrium soft matter is lacking at present.

Theoretical progress has recently been made for the case of many-body
systems governed by overdamped Brownian dynamics, encompassing a broad
spectrum of physical systems \cite{Hansen06}.  It has been
demonstrated that the dynamics of such systems can be described by a
unique time-dependent power functional $R_t[\rho,\Jv]$, where the
arguments are the space- and time-dependent one-body density
distribution, $\rho(\rv,t)$, and the one-body current distribution,
$\Jv(\rv,t)$, in the case of a simple substance
\cite{power,brader13noz}. Both these fields are microscopically sharp
and act as trial variables in a variational theory. The power
functional theory is regarded to be ``{\it important, [as it] provides
  (i) a rigorous framework for formulating dynamical treatments within
  the [density functional theory] formalism and (ii) a systematic
  means of deriving new approximations}''~\cite{EvansEtAl2016Preface}.

The physical time evolution is that which minimizes $R_t[\rho,\Jv]$ at
time~$t$ with respect to $\Jv(\rv,t)$, while keeping $\rho(\rv,t)$
fixed. Hence
\begin{align}
  \frac{\delta R_t[\rho,\Jv]}{\delta \Jv(\rv,t)}&=0
  \label{EQRtMinimizationPrinciple}
\end{align}
at the minimum of the functional.  Here the variation is performed at
fixed time $t$ with respect to the position-dependent current.  The
density distribution is then obtained from integrating the continuity
equation, $\partial \rho(\rv,t)/\partial t=-\nabla\cdot\Jv(\rv,t)$, in
time.  The power functional possesses units of energy per time, and
can be split according to
\begin{equation}
  R_t[\rho,\Jv] = P_t[\rho,\Jv] + \dot F[\rho] - X_t[\rho,\Jv],
  \label{EQRtSplitting}
\end{equation}
where $P_t[\rho,\Jv]$ accounts for the irreversible energy loss due to
dissipation, $\dot F[\rho]$ is the total time derivative of the
intrinsic (Helmholtz) free energy density functional
\cite{evans79,Hansen06}, and $X_t[\rho,\Jv]$ is the external power,
given by
\begin{align}
  X_t[\rho,\Jv] &= \int d\rv [\Jv(\rv,t)\cdot
    {\bf F}_{\rm ext}(\rv,t)  - \rho(\rv,t) \dot V_{\rm ext}(\rv,t)],
\end{align}
where $\dot V_{\rm ext}(\rv,t)$ is the partial time derivative of the
external potential $V_{\rm ext}(\rv,t)$, and ${\bf F}_{\rm
  ext}(\rv,t)$ is the external one-body force field, which in general
consists of a sum of a conservative contribution, $-\nabla V_{\rm
  ext}(\rv,t)$, and a further non-conservative term.  The power
dissipation is conveniently split into ideal and excess (above ideal)
contributions: $P_t[\rho,\Jv] = P_t^{\rm id}[\rho,\Jv]+P_t^{\rm
  exc}[\rho,\Jv]$, where $P_t^{\rm exc}[\rho,\Jv]$ is nontrivial and
arises from the internal interactions between the particles.  The
exact free power dissipation of the ideal gas is local in time and
space and given by
\begin{align}
  P_t^{\rm id}[\rho,\Jv] = \frac{\gamma}{2}\int d\rv 
  \frac{\Jv(\rv,t)^2}{\rho(\rv,t)},
  \label{EQPid}
\end{align}
where $\gamma$ is the friction constant of the Brownian particles
against the (implicit) solvent. This framework is formally exact and
goes beyond dynamical density functional theory
\cite{evans79,marinibettolomarconi99,archer04ddft}; the latter follows
from neglecting the excess dissipation, $P_t^{\rm exc}[\rho,\Jv]=0$.

In this Letter we apply the general framework of power functional
theory to treat phase coexistence of nonequilibrium steady states.
Such a state of $N$ particles in a volume $V$ at temperature $T$ is
characterized by a value of the total power functional taken at the
(local) minimum, $R_t^0(N,V,T)\equiv R_t[\rho^0,\Jv^0]$, where the
superscript~0 indicates a quantity at the minimum. We define the
chemical power derivative~$\nu$ and the (negative) volumetric power
derivative~$\pi$ via partial differentiation,
\begin{align}
  \nu = \left.\frac{\partial R_t^0}{\partial N}\right|_{V,T},\quad
  \pi &= -\left.\frac{\partial R_t^0}{\partial V}\right|_{N,T},
\end{align}
where $\nu$ and $\pi$ possess units of energy per time and pressure
per time, respectively.  In the limit of large $N$ and large $V$, the
specific free power per volume, $r_t(\rho_b)=R_t^0/V$, will depend
only on the (bulk) number density $\rho_b=N/V$; this implies the
identity $R_t^0 = - \pi V + \nu N$, which neglects possible surface
contributions. The simple relations $\nu=\partial r_t/\partial \rho_b$
and $\pi = -r_t + \rho_b \nu$ follow straightforwardly.  We shall
demonstrate below that the free power density, $r_t(\rho_b)$, is the
relevant physical quantity for analyzing phase behavior
out-of-equilibrium.

We assume that two coexisting nonequilibrium steady states, $A$ and
$B$, are characterized by particle number $N_A$ and $N_B$ and by
volume $V_A$ and $V_B$, respectively. The density in phase A (B) is
$\rho_A=N_A/V_A$ ($\rho_B=N_B/V_B$).  Hence, in a phase separated
state, the total power is a weighted sum,
\begin{align}
  R_t^0 = r_t(\rho_A) V_A + r_t(\rho_B) V_B,
  \label{EQRtSuperposition}
\end{align}
where the partial volumes of the two phases are
$V_A/V=(\rho_B-\rho_b)/(\rho_B-\rho_A)$ and
$V_B/V=(\rho_b-\rho_A)/(\rho_B-\rho_A)$, with
$\rho_A\leq\rho_b\leq\rho_B$.

The task of finding a global minimum of $R_t[\rho,\Jv]$ can now be
facilitated by a Maxwell common tangent construction on $r_t(\rho_b)$,
which implies the identities
\begin{align}
  r_t'(\rho_A) &= r_t'(\rho_B) =
  \frac{r_t(\rho_B)-r_t(\rho_A)}{\rho_B-\rho_A},
  \label{EQdoubleTangentInNonequilibrium}
\end{align}
where $r_t'(\rho_b)=\partial r_t(\rho_b)/\partial \rho_b$.  As a
consequence, both the chemical and the volumetric derivatives have the
same value in the coexisting phases:
\begin{align}
  \nu_A &= \nu_B, \quad  \pi_A = \pi_B,
  \label{EQcoexistenceEqualities}
\end{align}
and equality of temperature is trivial by construction. 

In order to illustrate this framework, we apply it to treat active
Brownian particles, which form a class of systems attracting much
current interest
\cite{CatesAndCompany1a,CatesAndCompany1b,CatesAndCompany2,phaseSeparationPapers}. We
consider spherical particles in $d$-dimensional space, with position
coordinates $\rv^N\equiv\{\rv_1\ldots\rv_N\}$ and (unit vector)
orientations $\boldsymbol\omega^N\equiv\{\boldsymbol\omega_1\ldots
\boldsymbol\omega_N\}$; here the orientational motion of each $\ov_i$,
where $i=1\ldots N$, is freely diffusive with orientational diffusion
constant $D_{\rm rot}$.  The swimming is due to an
orientation-dependent external force field, ${\bf F}_{\rm
  ext}(\ov_i)=\gamma s \ov_i$, which is nonconservative and does not
depend explicitly on $\rv$ and $t$; here $s$ is the speed for free swimming. We follow 
Refs. \cite{CatesAndCompany1a,CatesAndCompany1b} and use the
Weeks-Chandler-Andersen model, i.e.,~a Lennard-Jones pair potential,
which is cut and shifted at its minimum, such that the resulting
short-ranged pair force is continuous and purely repulsive. For
numerical convenience our Brownian dynamics (BD) simulations will be
performed in $d=2$.

Power functional theory provides a microscopic many-body expression
for $R_t^0$ \cite{power}. Omitting an irrelevant rotational
contribution, this is given (up to a constant $C$) by
\begin{align}
  R_t^0 = -\frac{\gamma}{2} \langle \sum_i \vel_i(t)^2 \rangle + C,
  \label{EQRtMicroscopic}
\end{align}
where the sum is over all particles and the angles denote a steady
state average.  To directly simulate the dissipated free power, we use
a discretized version of the instantaneous velocity
\cite{fortini14prl}, $\vel_i(t)=(\rv_i(t+\Delta t)-\rv_i(t-\Delta
t))/(2\Delta t)$, where $\Delta t$ is the time step of the standard
(Euler) computer simulation algorithm, where $\rv_i(t+\Delta
t)=\rv_i(t)+\gamma^{-1}\Delta t [-\nabla_i
  U(\rv^N)+\boldsymbol\xi_i(t)+{\bf F}_{\rm ext}(\ov_i(t))]$, with
$\boldsymbol\xi_i(t)$ being a Gaussian-distributed delta-correlated
noise term, with finite-difference, equal-time strength
$\langle\boldsymbol\xi_i(t)\cdot\boldsymbol\xi_j(t)\rangle=\delta_{ij}
k_BT d/(\gamma \Delta t)$; $C= Nk_BTd/(2\Delta t)$ is an irrelevant
constant, and $k_B$ is the Boltzmann constant. The external power is
given by
\begin{align}
  X_t &= \langle \sum_i \vel_i(t)\cdot 
  {\bf F}_{\rm ext}(\ov_i(t))\rangle, 
  \label{EQXtManyBody}
\end{align}
and we define the corresponding internal power, due to interparticle
interactions and Brownian forces, as
\begin{align}
  I_t = \langle \sum_i \vel_i(t)\cdot(-\nabla_i U(\rv^N)
  +{\boldsymbol \xi}_i(t)) \rangle.
\end{align}
This allows us to split \eqref{EQRtMicroscopic} into 
a sum of external and internal contributions,
\begin{align}
  R_t^0 = -I_t/2 - X_t/2.
  \label{EQRtAsInternalAndExternal}
\end{align}
By inserting \eqref{EQRtSplitting} into
\eqref{EQRtMinimizationPrinciple} and observing the structure
of~\eqref{EQPid}, it is straightforward to show that
\begin{align}
  I_t = -\dot F - 2 P_t^{\rm exc} 
  + \int d\rv d\ov \Jv(\rv,\ov,t)\cdot \left.
  \frac{\delta P_t^{\rm exc}[\rho,\Jv]}{\delta \Jv(\rv,\ov,t)}\right|_0,
  \label{EQItGeneralForm}
\end{align}
where the integrand is evaluated at the minimum, and we included the
argument $\ov$, treating the system effectively as a mixture of
different components \cite{brader14tpl}.

To sample \eqref{EQRtMicroscopic} efficiently in simulation, we
decompose the velocity as $\vel_i(t)=(\Delta \rv_i(t-\Delta
t)+\Delta \rv_i(t))/(2\Delta t)$, where $\Delta \rv_i(t) =
\rv_i(t+\Delta t)-\rv_i(t)$, given via the Euler algorithm as a sum of
three contributions, i.e., intrinsic, $\Delta \rv_i^{\rm
  int}(t)=-\Delta t\nabla_i U(\rv^N(t))$; random, $\Delta \rv_i^{\rm
  ran}(t)=\Delta t \boldsymbol\xi_i(t)$; and external, $\Delta
\rv_i^{\rm ext}(t)=\Delta t {\bf F}_{\rm ext}(\ov_i(t))$. Multiplying
out \eqref{EQRtMicroscopic} yields 36 contributions, of which we only
sample the three non-trivial types: $\langle\Delta\rv_i^{\rm
  int}(t)\cdot\Delta\rv_i^{\rm int}(t)\rangle$ and
$\langle\Delta\rv_i^{\rm int}(t)\cdot\Delta\rv_i^{\rm ext}(t)\rangle$
(where also similar contributions arise with one or both displaced
time arguments), as well as $\langle\Delta\rv_i^{\rm ran}(t-\Delta
t)\cdot\Delta \rv_i^{\rm int}(t)\rangle$. We use $N=1000$ and adjust $V$ in
order to control the density in the square simulation box with
periodic boundaries. The time step is chosen as $\Delta
t/\tau_0=10^{-5}$, where the timescale is
$\tau_0=\gamma\sigma^2/\epsilon$, with Lennard-Jones diameter $\sigma$
and energy scale $\epsilon$.  We allow the system to reach a steady
state in $10^7$ steps, and collect data for a further $10^8$ steps.
The rotational diffusion constant is set to $D_{\rm rot}=3k_BT/(\gamma \sigma^2)$,
and the external field strength is chosen as $s=24\sigma/\tau_0$.  The
Peclet number \cite{CatesAndCompany1a,CatesAndCompany1b}
 is ${\rm Pe}\equiv 3s/(D_{\rm rot}\sigma)= \gamma s \sigma / (k_BT)$.

\begin{figure}
  \includegraphics[width=0.9\columnwidth]{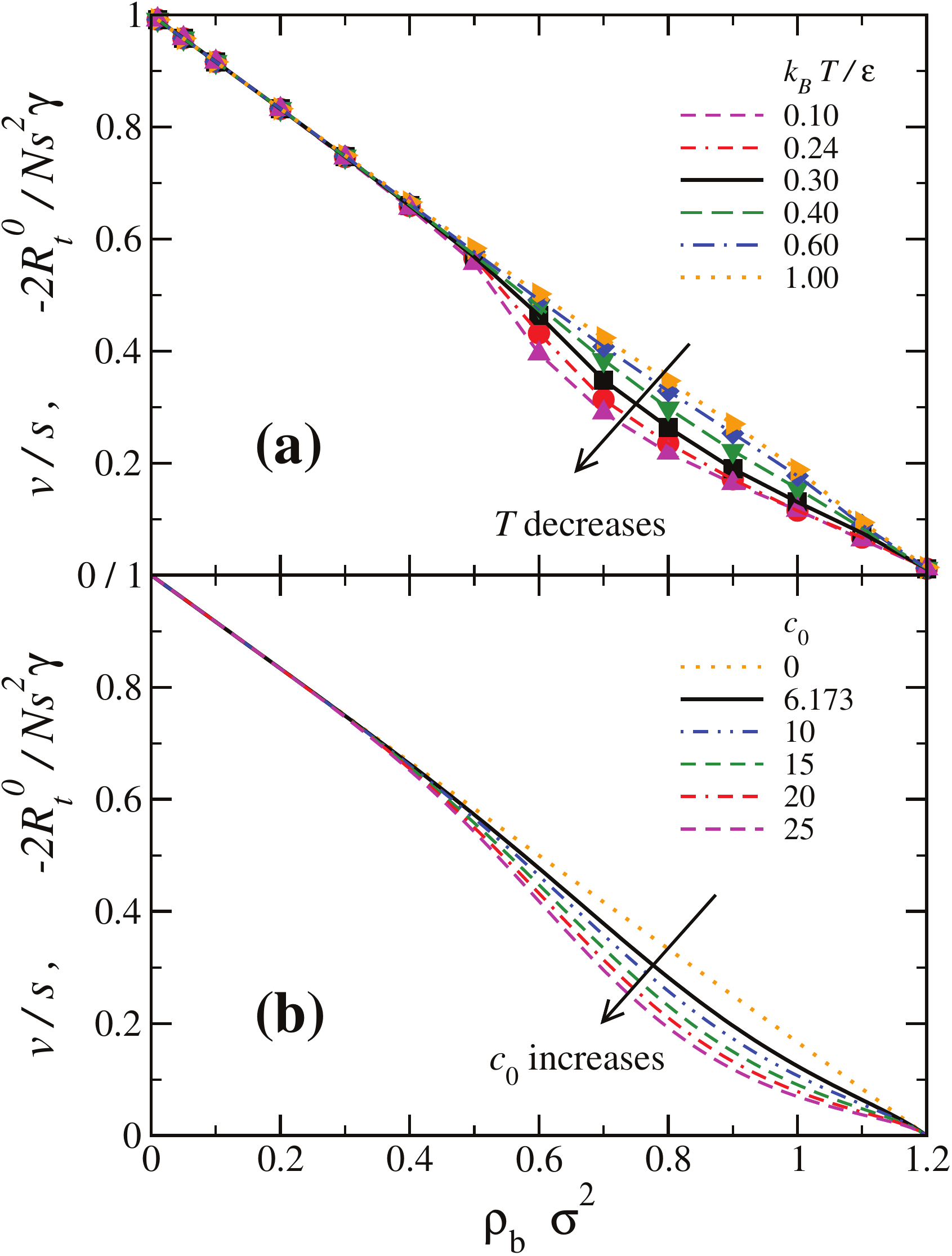}
\caption{(a) Scaled average forward swimming speed $v/s$ (symbols) and
  scaled free power $-2R_t^0/(Ns^2\gamma)$ per particle (lines),
  as obtained from BD computer simulations via
  Eqs.~\eqref{EQRtMicroscopic} and \eqref{EQXtManyBody}, respectively,
  for temperatures $k_B T/\epsilon =0.1$--1 (as indicated). (b)
  Theoretical results corresponding to (a), as given by
  Eqs.~\eqref{EQRtValueAtMinimum} and \eqref{EQvelocityMeanField},
  where $m=5$, $\rho_0\sigma^2=1.2$, and for values of $c_0=0$--25
  as indicated.}
\label{fig1}
\end{figure}

Figure \ref{fig1}(a) shows simulation results for $R_t^0$ and $X_t$, as
respectively given by \eqref{EQRtMicroscopic} and
\eqref{EQXtManyBody}, as a function of density.  Due to the simple
form of the external force, the external power \eqref{EQXtManyBody} is
trivially related to the (well-studied
\cite{CatesAndCompany1a,CatesAndCompany1b,CatesAndCompany2}) average
forward swimming speed~$v$ via $X_t = \gamma s v N$, where
$v=\langle\sum_i\vel_i(t)\cdot\ov_i\rangle/N$.  Remarkably, we find
that $R_t^0$ coincides with $-X_t/2$ within our numerical
precision. This implies that (i) the internal dissipation is
negligible, $I_t\approx 0$ (cf.~\eqref{EQRtAsInternalAndExternal}) and
(ii) that the value of the power functional for active particles is a
known quantity. We have systematically studied the variation with
temperature (as is analogous to varying Pe
\cite{CatesAndCompany1a,CatesAndCompany1b}). While hardly any effect
for low densities is observed, a dip develops for $\rho\sigma^2\gtrsim
0.5$, cf.~Fig.~\ref{fig1}a \cite{footnoteStenhammar}.

\begin{figure}
  \includegraphics[width=0.9\columnwidth]{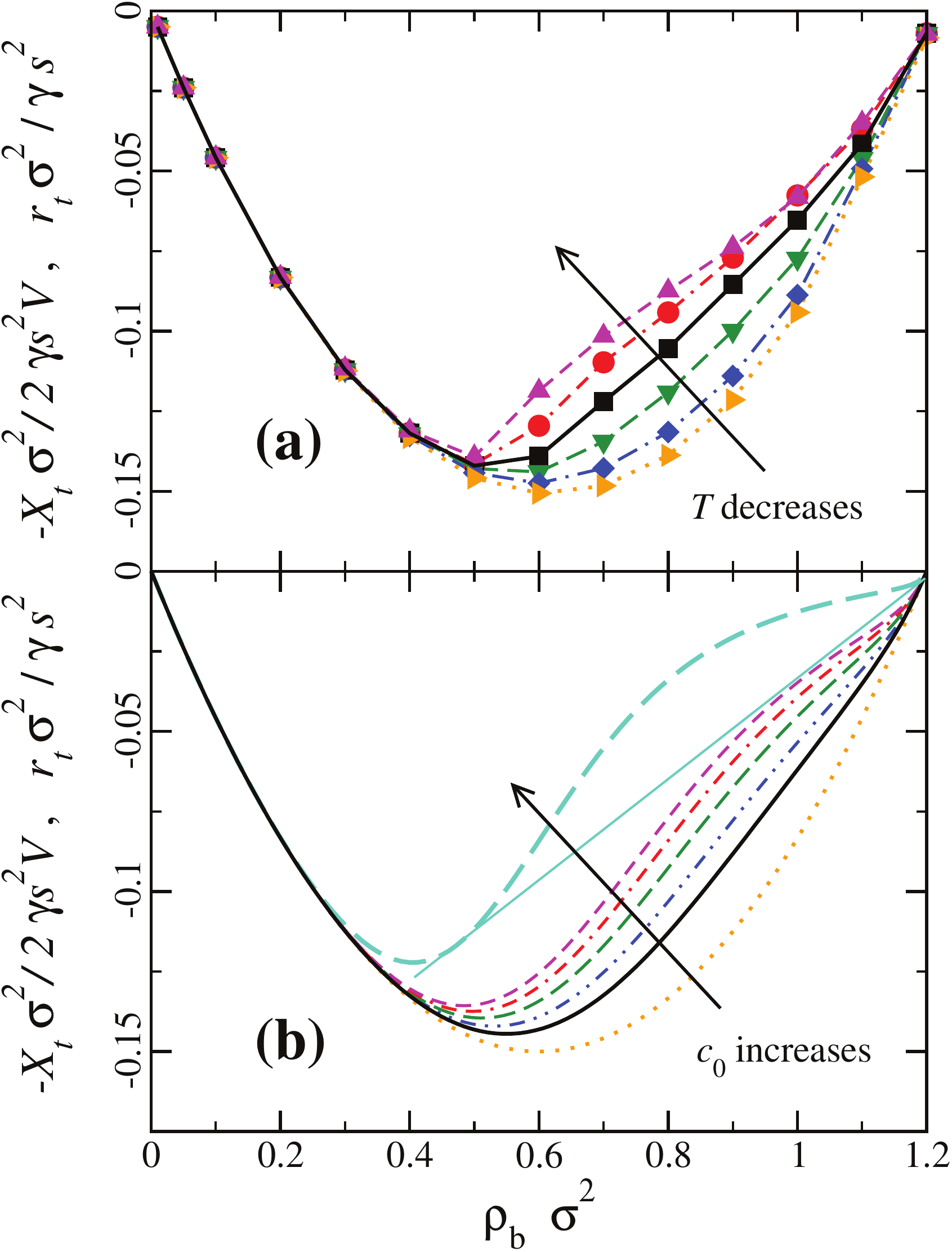}
\caption{The same as Fig.~\ref{fig1}, but scaled per volume $V$, rather
  than per particle $N$; the conversion factor is $-2V/\sigma^2$, such
  that $r_t\sigma^2/(\gamma s^2)$ and $-X_t\sigma^2/(2\gamma s^2V)$ are
  shown as a function of $\rho_b\sigma^2$.  The straight line in (b)
  indicates the double tangent for the case $c_0=100$; the black solid
  line indicates the result at the critical value of $c_0$.}
\label{fig2}
\end{figure}

\begin{figure}
  \includegraphics[width=0.9\columnwidth]{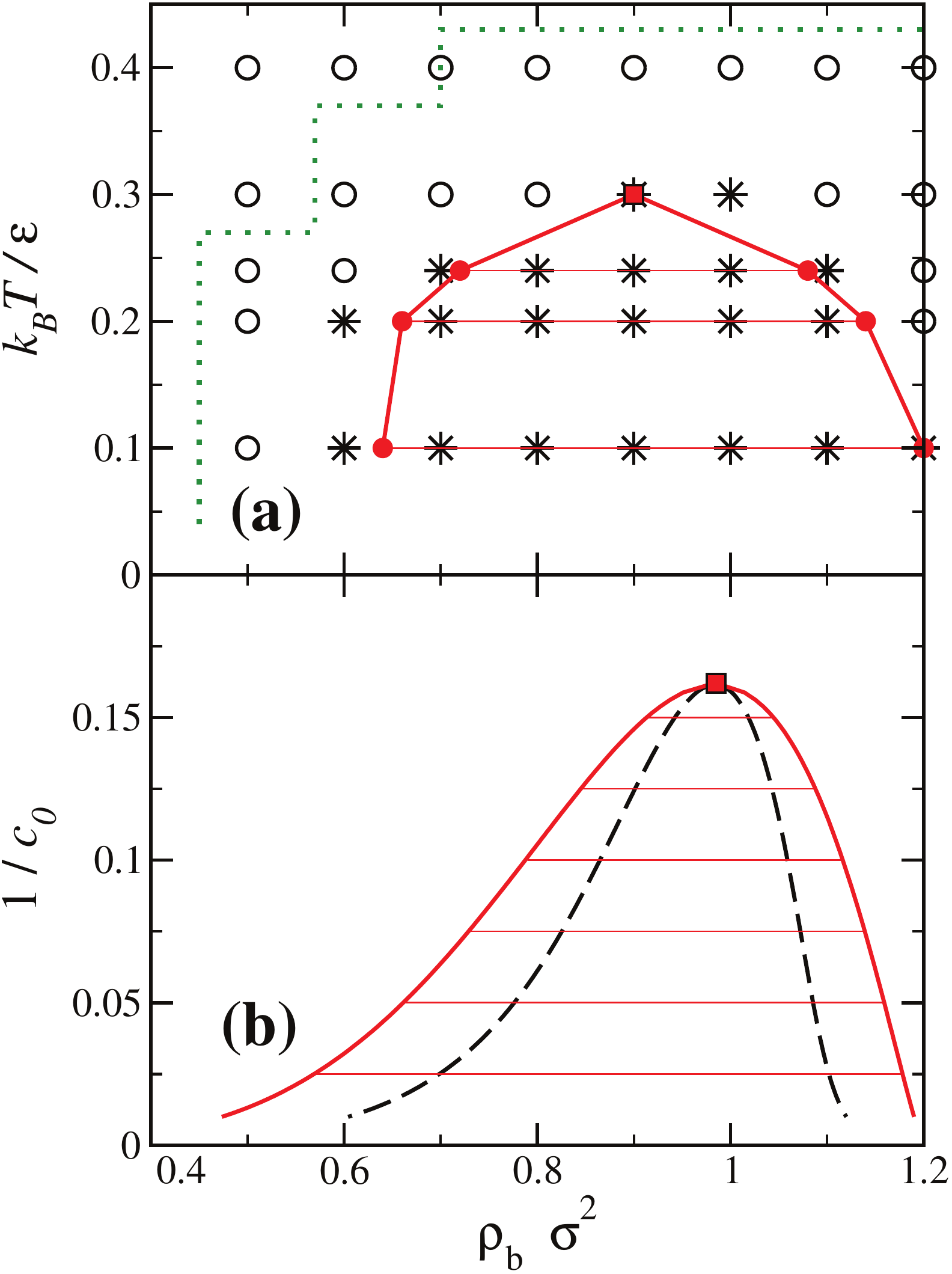}
\caption{(a) Phase diagram for active particles, obtained from
  simulations, as a function of scaled density $\rho_b\sigma^2$ and
  scaled temperature $k_BT/\epsilon$. Shown are the binodal (red solid
  line) obtained from double tangent construction (red filled
  symbols), horizontal tielines (thin red lines), and estimate for the
  critical point (red square). Also shown are single-phase (open
  symbols) and phase-separated (stars) states based on analysis of the
  decay of $g(r)$. The phase boundary of
  Ref.~\cite{CatesAndCompany1a,CatesAndCompany1b} is also shown (green
  dotted line).  (b) Same as (a), but obtained from power functional
  theory, and shown as a function $1/c_0$ instead of scaled
  temperature. The dashed line indicates the spinodal.}
\label{fig3}
\end{figure}

We next seek to develop a simple theoretical model to capture the key
features of the simulation data; the corresponding results shown in
Fig.~\ref{fig1}(b) will be discussed below. We assume $P_t^{\rm
  exc}[\rho,\Jv]$ to possess a simple Markovian, spatially nonlocal
form:
\begin{align}
  P_t^{\rm exc}[\rho,\Jv] &= \frac{\gamma}{2}
  \int \!d1 \!\!\int \!d2 \, \rho(1)\rho(2)
  \left(
  \frac{\J(1)}{\rho(1)}-\frac{\J(2)}{\rho(2)}
  \right)^2 \!M(1,2),
  \label{EQPexcApproximation}
\end{align}
where $1\equiv \rv,\ov$ and $2\equiv \rv',\ov'$. Here $M(1,2)$ is a
(dimensionless) correlation kernel that couples the particles at
points 1 and 2, similar to the mean-field form of the excess free
energy functional in equilibrium density functional theory
\cite{Hansen06,evans79}. Note that the term in brackets in
\eqref{EQPexcApproximation} is the (squared) velocity difference
between the two points.  We parameterize the current, which in general
depends on particle position $\rv$ and orientation $\ov$, as
$\Jv(\rv,\ov,t)= J_b \ov$, where the $J_b$ is a variational parameter
that determines the (homogeneous) bulk current in direction
$\ov$. This implies $v=J_b/\rho_b$.  Inserting into
\eqref{EQPexcApproximation} and observing the general structure
\eqref{EQRtSplitting}, we obtain
\begin{align}
  \frac{R_t}{\gamma V} &= \frac{J_b^2}{2\rho_b}
  +\frac{M_0}{2} J_b^2 - s J_b,
  \label{EQRtMeanField}
\end{align}
where the right hand side consists of a sum of contributions due to
ideal dissipation ($P_{\rm id}$), excess contribution to dissipation
($P_{\rm exc}$) and external power ($X_t$). The coefficient $M_0$ is
density-dependent and can be expressed as a moment of the correlation
kernel \cite{brader13noz} $M(1,2)$, as $M_0 = \int d\rv d\ov d\ov'
(\ov-\ov')^2 M(1,2)$, where due to symmetries $M(1,2)$ depends only on
the differences $\rv-\rv'$ and $\ov-\ov'$, and $M_0$ is hence
independent of $\rv'$.  Clearly, in steady states $\dot F[\rho]=0$.

The minimization principle \eqref{EQRtMinimizationPrinciple} implies
$\partial R_t/\partial J_b=0$ for \eqref{EQRtMeanField}, which yields
\begin{align}
  J_b &= s\rho_b/(1+M_0\rho_b).
  \label{EQJbGeneral}
\end{align}
Using \eqref{EQJbGeneral} in order to eliminate $M_0$ from
\eqref{EQRtMeanField} gives the value at the minimum
\begin{align}
    R_t^0  &= -\gamma s J_b V/2,
    \label{EQRtValueAtMinimum}
\end{align}
which implies that $R_t^0 = -X_t/2$, where here the external power is
$X_t=\gamma s J_b V$.  A detailed derivation will be given elsewhere.
The internal contribution $I_t=0$, as $\dot F=0$ in steady state, and
the additional contributions in \eqref{EQItGeneralForm} vanish for the
present form \eqref{EQPexcApproximation} of $P_{\rm exc}[\rho,\Jv]$,
which is quadratic in $\Jv(\rv,\ov,t)$.

We assume a simple analytical expression,
\begin{align}
  M_0 &= (\rho_0-\rho_b)^{-1} + c_0\rho_b^m/\rho_0^{m+1},
  \label{EQm0MeanField}
\end{align}
where $\rho_0$ is the jamming density at which the dynamics arrests,
$c_0\geq 0$ is a temperature-dependent dimensionless constant, and the
exponent $m>0$ is a measure for the number of particles that cause the
additional dissipation due to local cluster formation (second term in
\eqref{EQm0MeanField}). We expect the exponent $m$ to grow with $d$,
as clusters consist of an increasing number of particles upon
increasing $d$.  Furthermore, we expect $c_0$ to decrease to zero with
increasing temperature, as clusters are broken up by thermal motion.
We leave a microscopic derivation of $M_0$, e.g.\ starting from the
correlation kernel $M(1,2)$ (which is in principle accessible
e.g.\ via simulations \cite{schindler2016}) to future
work. Equation~\eqref{EQm0MeanField} can be interpreted as describing an
overall increase, and eventual divergence, of dissipation with density
plus a specific dissipation channel due to small groups of the order
of $m$ particles that block each other. Blocking is only relevant at
intermediate densities, high enough so that the $m$-th density order
contributes, but low enough in order to be not overwhelmed by the singularity.

Inserting \eqref{EQm0MeanField} into
\eqref{EQJbGeneral} yields
\begin{align}
  \frac{J_b}{s\rho_b}  &= \frac{1-x}{1+
  c_0x^{m+1}(1-x)},
  \label{EQvelocityMeanField}
\end{align}
where we have defined the scaled density $x=\rho_b/\rho_0$. In
case of high temperature, where $c_0\to 0$, this reduces to the simple
and well-known (see,
e.g.,~\cite{CatesAndCompany1a,CatesAndCompany1b,CatesAndCompany2})
linear (velocity) relationship $v/s \equiv J_b/(s\rho_b)=1-x$.  In
Fig.~\ref{fig1}(b), we show the theoretical results for the (scaled)
external and total free power per particle corresponding to the
simulation results in Fig.~\ref{fig1}(a). Clearly, despite the
simplicity of \eqref{EQm0MeanField} the theory reproduces the
simulation data very well.

As outlined above, in order to assess phase behavior, the relevant
quantity is the free power per volume $r_t$ (rather than per
particle), which we show in Fig.~\ref{fig2}, obtained from simulations
(Fig.~\ref{fig2}(a)) and theory (Fig.~\ref{fig2}(b)). For low temperatures
$k_BT/\epsilon=0.1,0.24$, the simulation data clearly show a
change in curvature, which we attribute to a first-order phase
transition in the finite system \cite{footnoteSystemSize}. (In an
infinite system, we expect no negative curvature to occur, and the
coexistence region to be characterized by a strictly linear variation
of $r_t$ with $\rho_b$.)  For $k_BT/\epsilon=0.3$ a quasilinear part
can be observed, which we interpret as being very close to a
nonequilibrium critical point.  The theoretical curve displays the
same type of behavior, which we attribute to the mean-field character
of the approximation \eqref{EQPexcApproximation}.  We can now apply
the general phase coexistence conditions
\eqref{EQdoubleTangentInNonequilibrium} and \eqref{EQcoexistenceEqualities}
to the active system.  A representative double tangent is shown in
Fig.~\ref{fig2}(b). The low-density (high-density) coexisting phase is
characterized by high (low) value of~$X_t$.

The phase diagram (cf.~Fig.~\ref{fig3}) displays two-phase coexistence
between a high-density and a low-density active fluid.  We find the
simulation results (Fig.~\ref{fig3}(a)) for the binodal obtained from
double tangent construction (on the results shown above in
Fig.~\ref{fig2}(a)) as a function of $k_BT/\epsilon$ to be consistent
with the behaviour of the tail ($5<r/\sigma<10$) of the radial pair
distribution function $g(r)$. A characteristic slow decay indicates
occurrence of phase separation (see, e.g.,\ \cite{farage2015}).  The
corresponding theoretical phase diagram is shown in Fig.~\ref{fig3}(b),
where we also display the spinodal, defined as the point(s) of
inflection of $r_t(\rho_b)$.  The phase separation vanishes upon
increasing $1/c_0$ at an upper nonequilibrium critical point.
Although we have not attempted to model the dependence of $1/c_0$ on
$T$ systematically, the agreement between simulation and theoretical
results is striking.  Our simulation results for the phase behaviour
underestimate the boundaries given by Stenhammar {\it et
al.}~\cite{CatesAndCompany1a,CatesAndCompany1b}; this is not surprising
given that these authors investigated significantly larger systems.
In simulations we have found only a slight decrease of the slope of
$v(\rho_b)$ for increasing $s$, and corresponding increase in the
jamming density, but with little effect on the phase separation
itself. This is consistent with the fact that $P_{\rm exc}[\rho,\Jv]$,
and hence, $c_0$ is an intrinisic quantity. The conditions for spinodal
and binodal both differ from the density ``where macroscopic MIPS [mobility-induced phase separation] is
initiated by spinodal decomposition'' \cite{CatesTailleurReview},
$v'/v = -1/\rho_b$, where $v'=dv(\rho_b)/d\rho_b$; this can be
rephrased as $d(\rho_b v)/d\rho_b=0$, implying, within $I_t=0$, that
$r_t'(\rho_b)=0$. This condition is quite different from the spinodal
within power functional theory, $r_t''(\rho_b)=0$, or equivalently $v''/v'=-2/\rho_b$.
Furthermore, for linear variation of $v$ with $\rho_b$, i.e.,\ $c_0=0$,
we find phase separation to be absent, in contrast to
Ref.~\cite{CatesTailleurReview}; cf.\ Eqs.(35)-(37) and Fig.~5
therein.

We have developed a general approach, based on power functional theory
\cite{power}, to treat coexistence between nonequilibrium steady
states in Brownian systems. Our theory is fundamentally different from
other approaches to active systems
(e.g.~\cite{phaseSeparationPapers,farage2015,furtherPapers}) which
were developed specifically for phase separation. We rather identify
a generating functional providing a unified, internally
self-consistent description of out-of-equilibrium states.  The free
power density plays a role in nonequilibrium systems analogous to that
of the free energy density in equilibrium, although it is an entirely
distinct physical quantity. 



\begin{thebibliography}{10}

\bibitem{weeks}
G.~L. Hunter and E. Weeks, Rep. Prog. Phys. {\bf 75}, 066501 (2012). 

\bibitem{schall}
V. Chikkadi, D. M. Miedema, M. T. Dang, B. Nienhuis, and P. Schall,
Phys. Rev. Lett. {\bf 113}, 208301 (2014).

\bibitem{dhont}
J.~K.~G. Dhont, Faraday Discuss. {\bf 123}, 157 (2003).

\bibitem{onuki}
A. Onuki, {\it Phase Transition Dynamics}, (Cambridge University Press, Cambridge, 2002).

\bibitem{phaseSeparationPapers}
Y. Fily, S. Henkes and M. C. Marchetti,
Soft Matter {\bf 10}, 2132 (2014);
J. Bialk{\'e}, H. L{\"o}wen and T. Speck, EPL {\bf 103}, 30008 (2013);

\bibitem{CatesTailleurReview}
M. E. Cates and J. Tailleur,
Annu. Rev. Condens. Matter Phys. {\bf 6}, 219 (2015).

\bibitem{Hansen06}
J.~P. Hansen and I.~R. McDonald, {\it Theory of Simple Liquids}, 4rd ed.
  (Academic Press, Amsterdam, 2013).

\bibitem{power}
M. Schmidt and J.~M. Brader, J. Chem. Phys. {\bf 138} 214101 (2013).   

\bibitem{brader13noz}
J.~M. Brader and M. Schmidt, J. Chem. Phys. {\bf 139}, 104108 (2013).

\bibitem{EvansEtAl2016Preface}
R. Evans {\it et al.}, J. Phys.: Condens. Matter {\bf 28}, 240401 (2016).

\bibitem{evans79}
R. Evans, Adv. Phys. {\bf 28},  143  (1979).

\bibitem{marinibettolomarconi99}
U. {Marini Bettolo Marconi} and P. Tarazona, J. Chem. Phys. {\bf 110},  8032
  (1999).

\bibitem{archer04ddft}
A.~J. Archer and R. Evans, J. Chem. Phys. {\bf 121},  4246  (2004).

\bibitem{CatesAndCompany1a}
J. Stenhammar, A. Tiribocchi, R. J. Allen, D. Marenduzzo, and M. E. Cates,
Phys. Rev. Lett. {\bf 111}, 145702 (2013);

\bibitem{CatesAndCompany1b}
J. Stenhammar, D. Marenduzzo, R. J. Allen and M. E. Cates,
Soft Matter {\bf 10}, 1489 (2014).

\bibitem{CatesAndCompany2}
A. P. Solon, J. Stenhammar, R. Wittkowski, M. Kardar, Y. Kafri,
M. E. Cates, and J. Tailleur, Phys. Rev. Lett. {\bf 114}, 198301 (2015);
A. P. Solon, Y. Fily, A. Baskaran, M. E. Cates, Y. Kafri, M. Kardar
and J. Tailleur, Nat. Phys. {\bf 11}, 673 (2015).

\bibitem{fortini14prl}
A. Fortini, D. de las Heras, J. M. Brader, M. Schmidt, 
Phys. Rev. Lett. {\bf 113}, 167801 (2014). 
The trajectory-based velocity is analogous to the operator
description of Refs.~\cite{power,brader13noz}.

\bibitem{brader14tpl}
J. M. Brader and M. Schmidt, J. Phys.: Condens. Matter {\bf 27}, 194106 (2015).

\bibitem{footnoteStenhammar} Nonlinear behaviour of simulation results
  for $v(\rho)$ has been reported before, see Fig.~S3 in the
  supplementary material of Ref.~\cite{CatesAndCompany1a}.

\bibitem{schindler2016}
T. Schindler and M. Schmidt, J. Chem. Phys. {\bf 145}, 064506 (2016).

\bibitem{footnoteSystemSize}
We expect the precise form of $r_t(\rho)$ and hence the value of
the phase coexistence densities to be affected by finite size effects.


\bibitem{farage2015} T. F. F. Farage, P. Krinninger, and J. M. Brader,
  Phys. Rev. E {\bf 91}, 042310 (2015).

\bibitem{furtherPapers}
I. Theurkauff {\it et al.}, Phys. Rev. Lett. {\bf 108}, 268303 (2012);
G. S. Redner {\it et al.}, Phys. Rev. Lett. {\bf 110}, 055701 (2013);
I. Buttinoni {\it et al.}, Phys.  Rev. Lett. {\bf 110}, 238301 (2013);
G. S. Redner {\it et al.}, arXiv:1603.01362 (2016);
D. Richard {\it et al.}, Soft Matter {\bf 12}, 5257 (2016).





\end{thebibliography}
\end{document}